\documentstyle[12pt]{article}
\newcommand{\bce}{\begin{center}}
\newcommand{\ece}{\end{center}}
\newcommand{\beq}{\begin{equation}}
\newcommand{\eeq}{\end{equation}}
\newcommand{\bea}{\vspace{0.25cm}\begin{eqnarray}}
\newcommand{\eea}{\end{eqnarray}}

\def\lsim{\mathrel{\rlap{\lower4pt\hbox{\hskip1pt$\sim$}}
    \raise1pt\hbox{$<$}}}         
\def\gsim{\mathrel{\rlap{\lower4pt\hbox{\hskip1pt$\sim$}}
    \raise1pt\hbox{$>$}}}         

\def\Pom{{\bf I\!P}}

\def\lsim{\mathrel{\rlap{\lower4pt\hbox{\hskip1pt$\sim$}}
    \raise1pt\hbox{$<$}}}         
\def\gsim{\mathrel{\rlap{\lower4pt\hbox{\hskip1pt$\sim$}}
    \raise1pt\hbox{$>$}}}         

\def\Pom{{\bf I\!P}}

  \advance\oddsidemargin  by -1in
  \advance\evensidemargin by -1in
  \advance\topmargin      by -0.5in
\def\lsim{\mathrel{\rlap{\lower4pt\hbox{\hskip1pt$\sim$}}
    \raise1pt\hbox{$<$}}}         
\def\gsim{\mathrel{\rlap{\lower4pt\hbox{\hskip1pt$\sim$}}
    \raise1pt\hbox{$>$}}}         

\def\Pom{{\bf I\!P}}
\def\beq{\begin{equation}}
\def\endeq{\end{equation}}
\def\arr{\begin{eqnarray}}
\def\endarr{\end{eqnarray}}
\makeindex

\begin{document}

\rightline{{\bf ITEP-PH-7/97}}
\vspace{1.5cm}
\begin{center}
{ \Large \bf
The BFKL-Regge Phenomenology \\
 of  Deep Inelastic Scattering
 \vspace{1.0cm}}

{\large\bf N.N. Nikolaev$^{a,b}$ B.G. Zakharov$^{b}$  and V.R.Zoller $^{c}$}\\
$^a$
Institut  f\"ur Kernphysik, Forschungszentrum J\"ulich,\\
D-52425 J\"ulich, Germany\\

$^b$ Landau Institute for Theoretical Physics,\\
GSP-1,117940, ul. Kosigyna 2, 117334 Moscow, Russia\\

$^c$Institute for Theoretical and Experimental Physics,\\
ul. B.Chermushkinskaya 25, 117218 Moscow, Russia\\
\vspace{0.5cm}

{\bf           Abstract}
\end{center}

We calculate the Regge trajectories
of the subleading BFKL  singularities and
 eigenfunctions
for the running BFKL pomeron in the color dipole representation.
 We obtain a viable
 BFKL-Regge  expansion of the
proton structure function $F_{2p}(x,Q^2)$
in terms of several rightmost
BFKL singularities.  We find large subleading contributions to
$F_{2p}(x,Q^2)$ in the HERA
 kinematical region
which   explains
the lack of  a predictive power of
GLDAP-extrapolations
of $F_{2p}(x,Q^2)$  to
a domain of small $x$. We point out the relation of our early finding
of the precocious BFKL asymptotics to the nodal structure of
subleading BFKL eigenfunctions.

\newpage



Extrapolation of the proton structure function
from the accessible region of $x$ and $Q^2$
to small $x$ remains the topical issue ever
 since the first SLAC experiments on deep inelastic
scattering (DIS). The GLDAP evolution \cite{GRIB72}
 has been a standard
instrument in these studies.
 It was soon realized that GLDAP
extrapolations are not unique, and equally good fits to
the large-$x$-data did invariably, and wildly, diverge with each other
and new experimental data,  when
extrapolated beyond the studied range of $x$.
For   instructive  comparison of
extrapolations of the pre-HERA fits to the proton structure
function with the new data from HERA
 down to $x \sim 10^{-5}$
 see \cite{F2HERA93}.
 The lack of predictive
power at small $x$ is not surprising: the GLDAP evolution
describes the future for large $Q^{2}$ starting with
assumptions on the parton densities
at past, $Q_{0}^{2}$. One needs to assume the whole function
of $x$ and different past defines a different future.

At very small $x$, the assumptions of the GLDAP evolution
break down and it is superseded by the $\log({1/x})$
BFKL evolution \cite{LIPAT76}. The BFKL
evolution is meant to predict the future at small $x$
from the past defined as a function of $Q^2$
at a sufficiently small $x_{0}$.
 At an asymptotically small $x$
the BFKL prediction, $F_{2}(x,Q^{2})=F_{\Pom}(Q^{2})
x^{-\Delta_{\Pom}}$, is unique modulo to the total normalization factor.
In the BFKL approach, the $x$-dependence of the structure function
at moderately small $x$ depends on the spectrum of eigenvalues
and on the eigenfunctions of the BFKL operator.
The purpose of the present study\footnote{The preliminary results
have been reported at the DIS'97 Workshop \cite{DIS97}} is an evaluation of
contributions from subleading BFKL singularities for the running
BFKL pomeron in the color dipole formulation \cite{PISMA1,NZZJETP,PLNZ1}.

The BFKL equation for the interaction cross section section
$\sigma(\xi,r)$ of the colour dipole $r$ with the target reads
(hereafter $\xi =\log(1/ x)$)
${\partial \sigma(\xi,r)/ \partial \xi} ={\cal K}\otimes
\sigma(\xi,r)$.
Here the kernel ${\cal K}$ is related to the wave function squared
of the color-singlet $q\bar{q}g$ state with the Weizs\"acker-Williams
 soft gluon. The gluon fields are calculated with the running QCD
coupling and
perturbative gluons are infrared regularized imposing
 a finite propagation radius
$R_c\simeq 0.2-0.3\,{\rm fm}$. The BFKL pole $\Pom_{n}$ corresponds
to the Regge-behaving solution of the BFKL equation,
\beq
\sigma_{n}(\xi,r)=\sigma_{n}(r)\exp(\Delta_{n}\xi)
=\sigma_{n}(r)\left({x_0\over x}\right)^{\Delta_{n}}\, ,
\label{eq:2.4}
\eeq
where the eigenfunction $\sigma_{n}(r)$ and the eigenvalue (intercept)
$\Delta_{n}$ are determined from
\beq
{\cal K}\otimes \sigma_{n}=\Delta_{n}\sigma_{n}(r)\, .
\label{eq:EVPROB}
\endeq
The behavior of eigenfunctions at $r\rightarrow 0$ and/or
$\alpha_{S}(r)\rightarrow 0$ has been found in \cite{NZZJETP,PLNZ1}:
\beq
\sigma_{n}(r)=
r^{2}\left[{1\over \alpha_{S}(r)}\right]^{\gamma_{n}-1}\,,
\,\, \gamma_{n}\Delta_{n}={4 \over 3}\, .
\label{eq:SMALLR}
\eeq
 Useful clues come from the experience
with solutions of the eigenvalue problem for the Schr\"odinger
equation. First, the leading eigenfunction $\sigma_{0}(r)$,
found numerically in \cite{PISMA1,NZZJETP,PLNZ1}, is node
free. Consequently, the subleading solutions with eigenfunctions
$\Delta_{n} < \Delta_{0}$ must have nodes. Second, the asymptotics
(\ref{eq:SMALLR}) must hold for all
eigenfunctions  in the region of
$r$ beyond all nodes. Third, because of the infrared regularization
$\sigma_{n}(r)$ saturate at
 $r\gsim R_{c}$. This suggests that for eigenfunctions with $n$
nodes one must try $\sigma_{n}(r)={\cal P}_{n}(z)\sigma_{0}(r)$
where
${\cal P}_{n}(z)$
is a polynomial of a suitably chosen variable $z\sim
[1/\alpha_S(Q^2)]^{\gamma}$.
 Then we apply
the variational principle and minimize the functional
\beq
\Phi(\sigma_{n})=
\int {dr \over r}
\left|{{\cal K}\otimes\sigma_n(r)-\Delta_{n}\sigma_n(r)
\over \sigma_{0}(r)}\right|^{2}
\label{eq:VARY}
\eeq
in the space of trial functions with the polynomial prefactor
${\cal P}_{n}(z)$. The success of such an unorthodox variational
principle for  excited states depends on the guessed trial
function, examples of astonishingly successful applications can be
found in \cite{NOVIK}.

With the so obtained eigenfunctions $\sigma_n(r)$ we can calculate
the contributions
$F_2^{(n)}(Q^2)$ to the proton structure function using the color
dipole factorization \cite{NZ91}
 and can perform
  the BFKL-Regge expansion
of the proton structure function
\beq
F_{2p}(x,Q^{2})=\sum_{n} A_{n} F^{( n)}(Q^{2})
\left({x_{0}\over x}\right)^{\Delta_{n}}\,.
\label{eq:REGGE}
\endeq

Our principal findings on solutions of the BFKL equation
 are as follows.

The running BFKL equation gives rise to a series of
 moving poles in the complex $j$-plane.
 The intercepts
$\Delta_n$ (Figure 1)
very closely, to better than $10\%$, follow the law
\beq
 \Delta_n= {\Delta_0\over (n+1)}\label{eq:DELTAN}
\eeq
suggested earlier by Lipatov from the quasiclassical consideration
\cite{LIPAT86}.The found eigenfunctions plotted as
${\cal E}_n(r)=\sigma_n(r)/r$ (Figure 2)
 to a crude
approximation are  similar to Lipatov's quasi-classical eigenfunctions,
which are ${\cal E}_n(r)\sim \cos[\phi(r)]$  for $n\gg 1$ \cite{LIPAT86}.
For a related numerical analysis of the running BFKL equation
 see \cite{ROSS}.
Within our specific infrared regularization the  node of $\sigma_{1}(r)$
is located at $r\simeq 0.05-0.06\,{\rm fm}$. With growing $n$, the location
of the first node
moves to a somewhat  larger $r$, and  the first
nodes  accumulate  at $r\sim 0.1-0.15\, {\rm fm}$.

The slope $\alpha_n^{\prime}$  of the Regge trajectory for the pole
$\Pom_n$ can be found using the technique of Ref.\cite{PISMA2,NZZslope}.
We find  $\alpha_n^{\prime}\simeq 0.07-0.06\,GeV^{-2}$ with
weak dependence on $n$.

The structure functions
$F_2^{(n)}(Q^2)$
for the $\Pom_n$ poles  are shown in Figure 3.
At large $Q^2$, far beyond the nodal region,
$F_2^{n}(x,Q^2)\propto
\left({x_0/ x}\right)^{\Delta_n}
\left[{1/ \alpha_S(Q^2)}\right]^{4/3\Delta_n}$.
Since the relevant variable is a power of the inverse gauge coupling,
which varies with $Q^2$ very slowly,
the nodes of $F_2^{(n)}(Q^2)$ are spaced by 2-3 orders of
magnitude in the $Q^2$-scale
 and only
the first  two of them  are  in the accessible range of $Q^2$.
The first nodes of $F_2^{(n)}(Q^2)$
are located at $Q^2\sim20-60\, GeV^2$. Hence, only the leading
structure function contributes significantly in this region.
 This explains the precocious  BFKL asymptotics for $Q^2\sim 60\,GeV^2$
 found earlier
 from the numerical solution of the color dipole running
 BFKL equation \cite{PLNZ2}.

An interesting finding is that the Born approximation for the
 dipole cross section, $\sigma_B(r)$, gives a very good quantitative
description of the proton structure function at $x_0\simeq 0.03$ \cite{PLNZ2}
The $r$-dependence of $\sigma_B(r)$ is quite different from that
 of the leading eigenfunction $\sigma_0(r)$, and expansion of
$\sigma_B(r)$ in BFKL eigenfunctions shows that the contribution
of subleading terms with $n>0$ makes up to $\sim60\%$ of $\sigma_B(r)$.
Consequently, the $Q^2$-dependence of the proton structure function at
$x=x_0$, and the subsequent $x$-dependence towards smaller $x$, is controlled
to a large extent by the subleading contributions $F_2^{(n)}(Q^2)$.

At small $x$ only $Q^2\lsim 10^3 \, GeV^2$ are accessible.
In this range the structure functions with $n\geq 3$
are hardly distinguishable. Besides, the splitting of intercepts
with $n\geq 3$ is much smaller than for $n=0,1,2$.
 Hence, the Regge expansion (\ref{eq:REGGE})
can be truncated at $n=2$ and  $F_2^{(2)}(Q^2)$
comprises contributions of all poles with $n\geq 2$.

The BFKL equation  allows one
to determine  the intercepts and structure functions $F_2^{(n)}(Q^2)$.
In the expansion  we put $A_0=1$. Then
the overall normalization of eigenfunctions, for instance,
$\sigma_n(r\gg R_c)=0.89\,{\rm mb}$, and $A_1=0.39$ and $A_2=0.33$ are two
  adjustable
parameters which are fixed from
 the boundary condition at $x=x_0$.
With the  proper account of the  valence \cite{GLUCK}
and non-perturbative \cite{PLNZ2} corrections to (\ref{eq:REGGE}) we arrive at
the three-pole-approximation
 which provides a viable description of
the experimental data \cite{DATH1}
in a wide kinematical range (Figure 4).
Notice that in the pre-nodal region of
$Q^2\lsim 20 \, GeV^2$  the leading and subleading
 structure functions
are very close in shape and the experimental data in such a limited range
of $Q^2$ are absolutely insufficient for the determination of $A_n$, which
explains the failure of the early GLDAP fits.

The effective pomeron intercept
\begin{equation}
\Delta_{eff}=-{\partial \log F_{2p}(x,Q^2)\over \partial \log x}
 \label{eq:DELEFF}
\end{equation}
gives an idea of  the role of the subleading singularities.
The intercept $\Delta_{eff}$
  calculated with  the experimental kinematic constraints
 is much smaller than $\Delta_0=0.4$
which is expected to dominate asymptotically.
 The agreement of our numerical estimates
 with the $H1$ determination \cite{HERA97} (Figure 5)
is quite satisfactory.

\vspace{0.5cm}

B.G. Zakharov and V.R. Zoller thank J. Speth for the hospitality in
the Institut f\"ur Kernphysik, Forshungszentrum J\"ulich, where
this work was finished. This work was partly supported by the DFG
grant 436-RUS-17/13/97 and INTAS
grant 93-239ext.

\newpage
{\bf Figure captions}\\

\vspace{1cm}

Fig.1  BFKL eigenvalues.\\

Fig.2  BFKL eigenfunctions.\\

Fig.3  Modulus of BFKL structure functions.\\

Fig.4  Three-pole approximation vs. the data of H1, ZEUS and E665.\\

Fig.5  Effective intercept vs. H1 determination \cite{HERA97}.\\


\newpage

\begin{thebibliography}{99}




\bibitem{GRIB72}Gribov V.N., and Lipatov L.N., {\it Sov. J. Nucl.\  Phys.}\
 {\bf 15}  (1972) 438; Dokshitzer Yu.L., {\it Sov.\  Phys.\ JETP}\
 {\bf 46}  (1977) 641; Altarelli G., and Parisi G., {\it Nucl.\  Phys.}\
 {\bf B126} (1977) 297.
\bibitem{F2HERA93}H1 Collab., Abt I. et al., {\it  Nucl.\  Phys.} {\bf B44}
(1993) 515.
\bibitem{LIPAT76}Fadin V.S., Kuraev E.A.,and Lipatov L.N.,
{\it Sov.\  Phys.\ JETP}\
 {\bf 44} (1976) 443; ibid {\bf 45} (1977) 199;
Yu.Yu.Balitsky and L.N.Lipatov,
{\sl Sov. J. Nucl. Phys.} {\bf 28} (1978) 822.



\bibitem{DIS97} Zoller V.R.,  {\it Talk at 5th International Workshop
on Deep Inelastic Scattering and QCD (DIS'97), Chicago, April 1997},
Report No. KFA-IKP(TH)-97-12.


\bibitem{PISMA1}Nikolaev N.N., Zakharov B.G., and Zoller V.R.,
{\it  JETP\ Letters}\
 {\bf 59} (1994) 8.
\bibitem{NZZJETP}Nikolaev N.N., Zakharov B.G., and Zoller V.R.,
{\it JETP} {\bf 105}
(1994) 1498.

\bibitem{PLNZ1}Nikolaev N.N., Zakharov B.G., {\it Phys.\ Letters}\
 {\bf B327}  (1994) 157.

\bibitem{NOVIK}Karl G., and Novikov V. {\it  Phys.\ Rev}\
 {\bf D51}  (1995) 5069.

\bibitem{NZ91}
N.N.~Nikolaev and B.G.~Zakharov, {\it Z. Phys.} {\bf C49} (1991) 607;
{\it Z. Phys.} {\bf C53} (1992) 331.

\bibitem{LIPAT86} Lipatov L.N., {\it Sov.\  Phys.\ JETP}\ {\bf 63}   (1986) 904
\bibitem{ROSS} Hancock R.E. and Ross D.A. {\it Nucl.\ Phys.}\ {\bf B383} (1992)
575



\bibitem{PISMA2}Nikolaev N.N., Zakharov B.G., and Zoller V.R.,
{\it  JETP\ Letters}\
 {\bf 60} (1994) 678.
\bibitem{NZZslope}Nikolaev N.N., Zakharov B.G., and Zoller V.R.,
{\it  Phys.\ Letters}\
 {\bf B366}  (1996) 337.


\bibitem{PLNZ2}Nikolaev N.N., and Zakharov B.G., {\it Phys.\ Letters}\
 {\bf B333}  (1994) 250.

\bibitem{GLUCK}Gl\"uck M., Reya E., and Vogt A,    {\it Z.\ Phys.}\
 {\bf C67}  (1995) 433.

\bibitem{DATH1} H1 Collab., Ahmed T. et al.,  {\it Nucl.\ Phys.}\
 {\bf B439} (1995) 471;  ZEUS Collab., Derrick M. et al.,   {\it Z.\ Phys.}\
 {\bf C69}  (1996) 607;  E665 Collab., Adams M.R. et al.,  {\it Phys.\ Rev.}\
 {\bf D54}  (1996) 3006.

\bibitem{HERA97} H1 Collab., Aid S. et al.,  {\it Nucl.\ Phys.}\
 {\bf B470} (1996) 3.



\end{thebibliography}
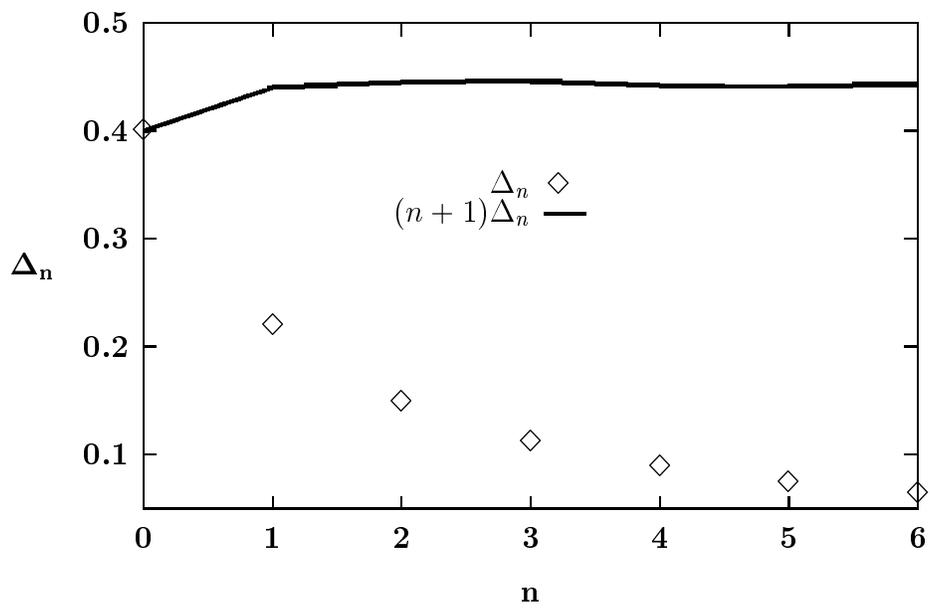
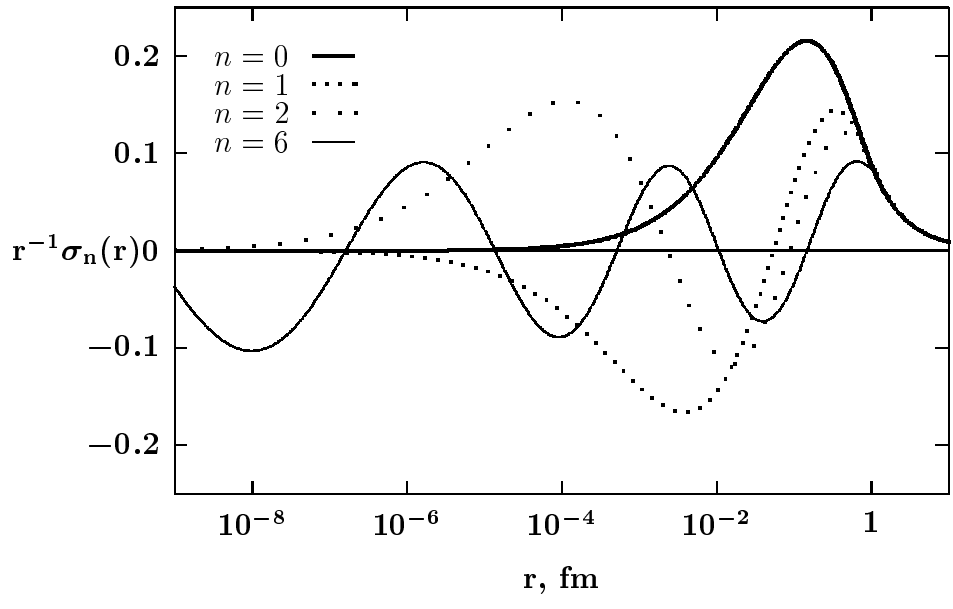
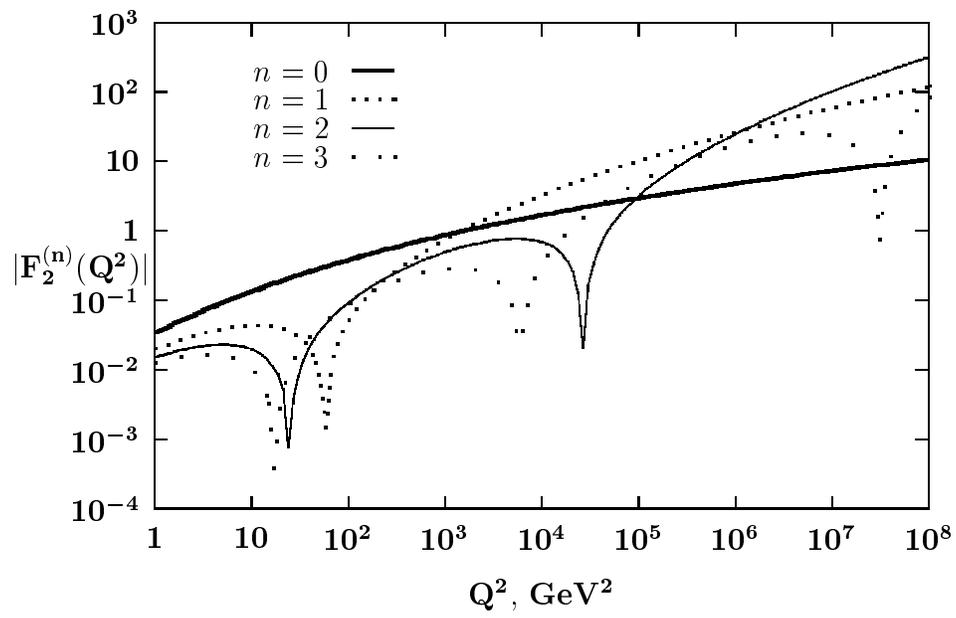
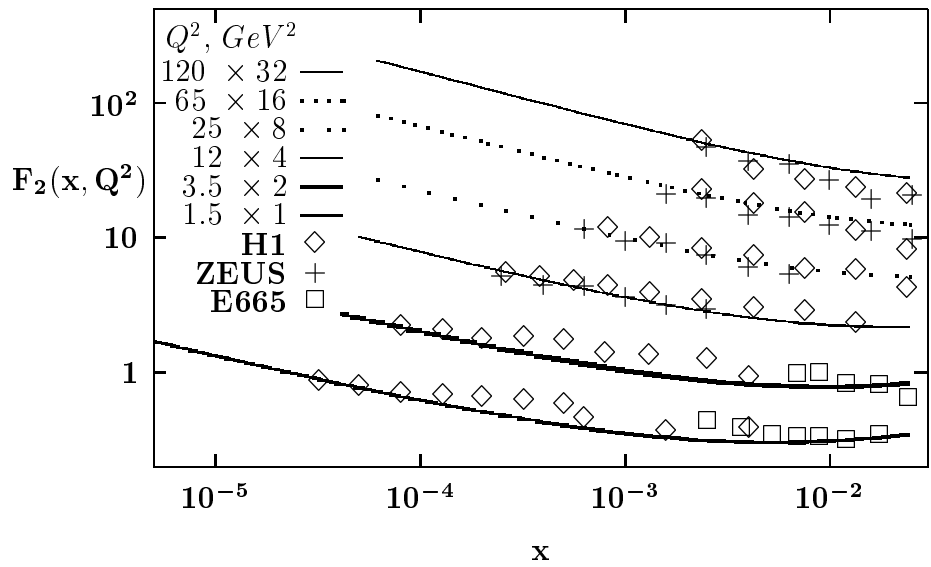
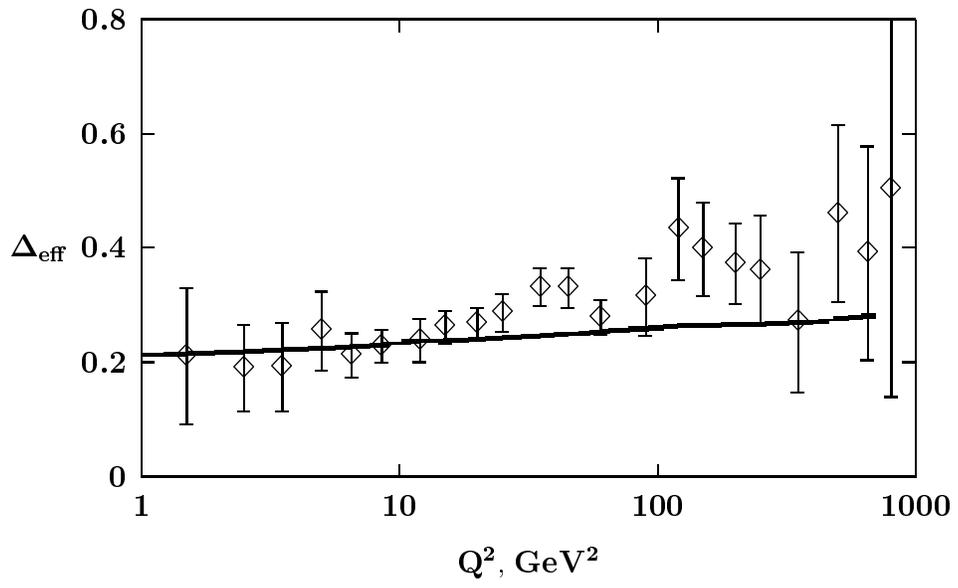
\end{document}